\documentclass[12pt]{iopart}
\usepackage[english]{babel}

\usepackage[letterpaper,top=2cm,bottom=2cm,left=3cm,right=3cm,marginparwidth=1.75cm]{geometry}

\expandafter\let\csname equation*\endcsname\relax
\expandafter\let\csname endequation*\endcsname\relax

\usepackage{amsmath}
\usepackage{amsfonts}
\usepackage{graphicx}
\usepackage{subcaption}
\usepackage{enumitem}
\usepackage{float}
\usepackage{comment}
\usepackage[sorting=none]{biblatex}
\usepackage[colorlinks=true, allcolors=black]{hyperref}

\begin{document}
\title[Theory-informed neural networks for particle physics]{Theory-informed neural networks \\ for particle physics}

\author{Barry M. Dillon$^{\text{1}}$, Michael Spannowsky$^{\text{2}}$}

\address{$^{\text{1}}$ISRC, Ulster University, Derry, BT48 7JL, Northern Ireland}
\address{$^{\text{2}}$IPPP, Durham University, Durham, DH1 3LE, U.K.}

\ead{b.dillon@ulster.ac.uk, michael.spannowsky@durham.ac.uk}

\vspace{1pt}

\begin{abstract}
\noindent We present a theory-informed reinforcement-learning framework that recasts the combinatorial assignment of final-state particles in hadron collider events as a Markov decision process. 
A transformer-based Deep Q-Network, rewarded at each step by the logarithmic change in the tree-level matrix element,
learns to map final-state particles to partons.
Because the reward derives solely from first-principles theory, the resulting policy is label-free and fully interpretable, allowing every reconstructed particle to be traced to a definite partonic origin. 
The method is validated on event reconstruction for $t\bar{t}$, $t\bar{t}W$, and $t\bar{t}t\bar{t}$ processes at the Large Hadron Collider.
The method maintains robust performance across all processes, demonstrating its scaling with increasing combinatorial complexity.
We demonstrate how this method can be used to build a theory-informed classifier for effective discrimination of longitudinal $W^{+}W^{-}$ pairs, and show that we can construct theory-informed anomaly-detection tools using background process matrix elements.
Building on theoretical calculations, this method offers a transparent alternative to black-box classifiers.
Being built on the matrix element, the classification and anomaly scores naturally respect all physical symmetries and are much less susceptible to the implicit biases common to other methods.
Thus, it provides a framework for precision measurements, hypothesis testing, and anomaly searches at the High-Luminosity LHC.
\end{abstract}
\vspace{4pt}

\section{Introduction}
Reconstructing the microscopic dynamics underlying complex final states at high-energy colliders is a central challenge in experimental particle physics. Events at the Large Hadron Collider (LHC) often produce multiple jets, leptons, and missing energy signatures, many of which stem from the decays of short-lived particles such as top quarks, $W/Z$ bosons, or hypothetical new states. Determining how these observed particles are assigned to hypothesised decay chains—identifying which jet originated from which parton in the matrix element is essential for precision Standard Model (SM) measurements and searches for physics beyond the Standard Model (BSM). Yet, this problem is inherently combinatorial: as the number of final-state objects increases, the number of possible assignments grows factorially, rendering brute-force likelihood evaluations intractable.

Traditionally, this challenge was approached using matrix element methods \cite{Kondo:1988yd,Abazov:2004cs,Soper:2011cr,Soper:2012pb,Soper:2014rya} or multivariate analyses based on kinematic observables \cite{Plehn:2001nj,Barr:2007hy,Butterworth:2008iy,Plehn:2009rk}. More recently, machine-learning has been employed to automate and accelerate the decision-making process, particularly through supervised classification using neural networks \cite{Almeida:2015jua,deOliveira:2015xxd,Baldi:2016fzo}. However, these methods have notable limitations. Supervised networks require labelled data, often depend heavily on simulation, and yield black-box classifiers that obscure the physical reasoning behind their outputs. Moreover, they are not easily interpretable, and their performance is typically optimised for specific tasks, such as signal-background separation, rather than for building an interpretable and physically meaningful event topology.
Developing interpretable machine-learning approaches is key to understanding how decisions are made in the analysis and should give us a better understanding of the systematic uncertainties in the analysis \cite{Faucett:2020vbu}.

As formalised by the Neyman-Pearson lemma \cite{1933RSPTA.231..289N}, an idealised classification scheme would use the likelihood ratio between competing hypotheses to produce the most powerful discriminator at a fixed false-positive rate. In principle, the squared matrix element provides this optimal test statistic, encoding all available information about spins, masses, and angular correlations. Yet, integrating this knowledge into trainable, scalable learning algorithms remains an unsolved task.

This paper introduces theory-informed reinforcement-learning (RL) as that bridge.  The assignment of final-state particles is framed as a Markov decision process: at each step, an agent chooses a swap or “do-nothing” action that reshuffles the event graph, receives a reward equal to the logarithmic change in the exact tree-level matrix element, and iteratively builds the topology that maximises the total discounted reward \cite{Ngairangbam:2023cps}. Deep Q-Networks \cite{mnih2013playing,mnih2015human} with permutation-invariant transformer backbones serve as policy and target networks, with a positional encoding for the mapping between final state particles and partons,
so that the agent learns kinematics and the hierarchical pattern of decays.  
Extensive training on simulated parton-level samples shows that the agent rapidly converges to the correct top quark decay chain, discriminates longitudinally polarised W pairs, and flags anomalous events without being told what the “right answer” looks like.

By intertwining exact quantum-field-theory calculations with the exploration capabilities of RL, the approach delivers three concrete benefits.  First, it exploits the full information content of the matrix element, which automatically encodes spin correlations and interference effects that are hard to approximate with handcrafted observables.  Second, the learned policy produces an explicit event graph, giving experimentalists transparent access to which particle ends up in which decay slot.  Third, because the reward is derived from theory rather than labelled data, the same agent can be reused for tagging, anomaly-detection, or hypothesis testing with minimal retraining.  The study, therefore, charts a path toward interpretable, theory-consistent machine-learning pipelines for the High-Luminosity LHC and beyond\footnote{Code for the project will be maintained at \url{https://github.com/bmdillon/theory-informed-neural-networks}}.

In Section~\ref{sec:dql}, we define the problem of parton assignment and describe its formulation as a reinforcement-learning task. Section~\ref{sec:tinnq} outlines the design of the agent, including the network architecture, training strategy, and the physics-based reward function derived from matrix elements. 
Section~\ref{sec:evreco} presents numerical experiments demonstrating the agent’s ability to reconstruct top quark decay chains, distinguish $W$ boson polarisation, and detect anomalous event structures. 
Finally, in Section~\ref{sec:ticl} we discuss broader implications, limitations, and future directions, including the potential to apply this framework to detector-level data and more complex decay topologies. Finally, we offer a summary and conclusions in Section~\ref{sec:summary}.

\section{Deep Q-Learning}
\label{sec:dql}
Reinforcement-learning \cite{watkins1992,SuttonBarto2018} is based on the concept of an agent interacting with an environment to solve a problem over a series of fixed steps.
At any particular step $t$, we write the state as $s_t\!\in \mathcal{S}$ where $\mathcal{S}$ is the set of all possible states that we can have.
At each step, the agent observes the state $s_t$ and decides on an action to take.
The action at step $t$ is denoted $a_t\!\in\! \mathcal{A}$ where $\mathcal{A}$ is the set of all possible actions that the agent can take.
When the agent takes an action, it changes the state from $s_t\!\rightarrow\!s_{t+1}$.
The agent then needs to receive some feedback on how good this change of state was; we call this the reward and write it as $r_t\!=\!f_r(s_t,a_t)$.
The agent uses the reward to learn how to choose better actions; the reward is arguably the most important aspect of RL.
In this paper, the agent receives a reward at each step, but in some cases, it is necessary to have the agent receive sparse rewards.
In the extreme case, the agent will only receive a reward on the final step.

The goal in RL is to learn a policy function $\pi(s)\!:\!\mathcal{S}\!\rightarrow\!\mathcal{A}$.
The agent uses this policy function to choose the next action to take.
The policy function $\pi(s)$ only considers the state at the current step when deciding what action to take.
This assumes that the problem can be solved as a Markov Decision Process (MDP), where the current reward and the next state can be computed solely in terms of the current state and the current action taken.
The optimal policy function should try to maximise the expected rewards $r_t$ over all future steps, not just the next step.
The total reward over all steps is written as $\sum r_t\!=\!r_0\!+\!r_1\!+\!\ldots\!+\!r_T$.
However, it can be difficult to learn a policy when considering all future rewards, and so we instead use the discounted reward at each step $t$:
\begin{equation}
    R_t = \sum_{k=0}^{T-t}\gamma^k r_{k+t}
\end{equation}
where $\gamma$ is the discount factor that determines how much future rewards are valued, and $T$ is the maximum number of steps to be taken (can be $\infty$).
Given some state $s_t$ and some policy function $\pi(s)$, we define the expected return $V^{\pi}\!:\!\mathcal{S}\!\rightarrow\!\mathbb{R}$ as
\begin{equation}
    \label{eq:valuefunc}
    V^{\pi}(s)=\mathbb{E}\left[R_t\Big| s_t\!=\!s,\pi\right].
\end{equation}
Policy functions can be defined to be either stochastic or deterministic.
In this paper, we employ a deterministic policy function, which means that for every state $s_t$, we obtain a definite action to take.

There are several ways to obtain a policy function that maximises Eq.~\ref{eq:valuefunc}.
Here we will take a value-based approach called Deep Q-Learning.
With Q-learning we define a state-action function $Q:\mathcal{S}\times\mathcal{A}\rightarrow \mathbb{R}$ that returns the expected discounted reward $R_t$ given a state $s_t$ and an action $a_t$,
\begin{equation}
    \label{eq:qfunc}
    Q(s,a)=\mathbb{E}\left[R_t\Big| s_t\!=\!s,a_t\!=\!a\right].
\end{equation}
In traditional Q-learning, this function is defined as a look-up table where the rewards are recorded for every state-action pair.
For complex problems, the number of pairs can be exponentially large.
In this case, we can approximate $Q(s,a)$ using deep neural networks, i.e. Deep Q-Learning.
If we have an accurate approximation to the $Q(s,a)$ function, we also have an accurate approximation to the optimal policy function
\begin{equation}
    \pi^{*}(s) = \underset{a\in\mathcal{A}}{\max}~Q^{*}(s,a).
\end{equation}
The challenge is then to learn an accurate approximation to $Q(s,a)$.

\subsection*{Learning the Q-function}

The expression for $Q(s,a)$ in Eq.~\ref{eq:qfunc} can be rewritten in a recursive form called the Bellman equation:
\begin{equation}
    \label{eq:bellman}
    Q(s,a)=\mathbb{E}\left[ f_r(s,a)+\gamma \sum_{s^\prime} T(s^\prime|s,a)~\underset{a^\prime}{\max}~Q(s^\prime,a^\prime) \right]
\end{equation}
where $(s,a)$ is the current states and action taken, while $(s',a')$ are the future states and actions taken.
The function $f_r(s,a)$ is used to compute the rewards, and $T(s^\prime|s,a)$ is the transition function that determines the next state given the current state and current action chosen.
The Bellman equation plays the role of the loss function in Deep Q-Learning, where we use a policy network to approximate $Q(s,a)$ and a target network to approximate $Q(s^\prime,a^\prime)$.
Both of these networks have the same architecture, with outputs of dimension $d_{\mathcal{A}}$, the dimension of the action space.

\begin{figure}[h]
    \centering
    \includegraphics[width=1.0\linewidth]{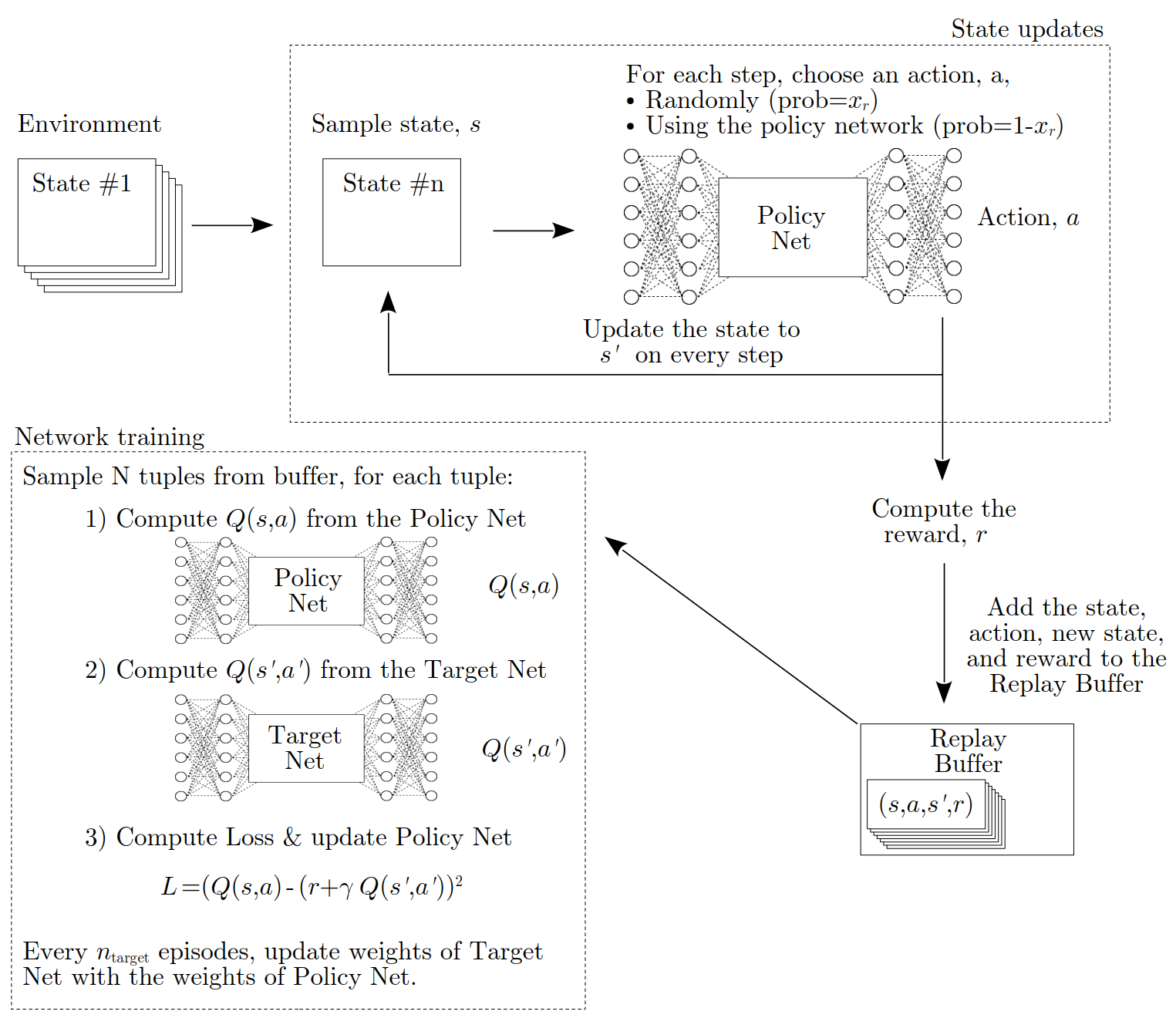}
    \caption{Outline of the general training procedure in Deep Q-Learning.}
    \label{fig:dql-general-schema}
\end{figure}

The training is organised in a series of episodes.
In each episode, the agent samples an initial state from the environment and uses the policy function to determine the next states.
The agent continues this process until the maximum number of steps, $T$, is reached.
To encourage the agent to explore different choices of actions, we introduce some randomness into the action selection.
On every step, we compute a random number in $[0,1]$ and if this is less than a parameter $x_r$, then we choose an action at random instead of using the agent's policy function.
A replay buffer is used to store the agent's experiences, i.e. the states, actions, and rewards that it encounters while moving through the steps.
After each episode, we sample data from the replay buffer and, using the Bellman relation, we update the weights of the policy network to improve the Q-value predictions.
This uses both the rewards and states sampled from the replay buffer, and the target network approximation to $Q(s^\prime,a^\prime)$.
Every $n_{\text{target}}$ episodes, we update the weights of the target network to match those of the policy network.
In some sense, the policy network is trying to converge to a moving target, but with $n_{\text{target}}\!\gg\!1$ we give the policy network time to converge slightly before changing the target network.
The outline of the training procedure is shown in Fig. \ref{fig:dql-general-schema}, and outlined here for clarity:
\begin{enumerate}[label*=\arabic*.]
    \item Define two neural-networks with the same architecture, $F_{\text{policy}}(s)$ and $F_{\text{target}}(s)$, and initialise with the same weights
    \item For each episode
        \begin{enumerate}[label*=\arabic*.]
            \item Select a state $s$ from the environment
            \item for each step
                \begin{itemize}
                \item[-] if rand$()\!<\!x_{\text{r}}$ then choose action $a$ at random
                \item[-] else, choose action $a$ using $F_{\text{policy}}(s)$
                \item[-] compute the reward $r$ for the action and the next state $s^\prime$
                \item[-] add $(r,s,s^\prime)$ to a replay buffer
                \end{itemize}
            \item Sample batch of $N$ $(r,s,s^\prime)$ from the replay buffer
            \item For each item in the batch
            \begin{itemize}
                \item[-] compute $Q(s,a)$ using $F_{\text{policy}}(s)$
                \item[-] compute $Q(s^\prime,a^\prime)$ using $F_{\text{target}}(s^\prime)$
                \item[-] compute mean-squared-error (MSE) $L= \left[Q(s,a) - \left(r+\gamma Q(s^\prime,a^\prime)\right)\right]^2$
                \item[-] update the weights of $F_{\text{policy}}(s)$ to minimize $L$
            \end{itemize}
            \item every $n_{\text{target}}$ episodes update the weights of $F_{\text{target}}(s)$ with those of $F_{\text{policy}}(s)$
        \end{enumerate} 
\end{enumerate}
As long as we train with enough episodes and allow the policy network enough episodes to converge before updating the target network weights, then the result at the end should be a policy network that can accurately approximate $Q(s,a)$ and the policy function $\pi(s)$.

\section{Theory-informed Q-Networks}
\label{sec:tinnq}

In this section, we will outline a method for mapping final-state particles measured in a collider to partons in a given process defined by theory.
The aim is to find the optimal mapping from these final-state particles to the partons of a given process, which means assigning the correct final-state particle to the correct parton in the process.
We define the optimal assignment as the one which maximises the matrix element for that process, although this is not always the `true' assignment.
We focus on $2\!\rightarrow\!N$ processes, either gluon- or quark-initiated, and to develop and demonstrate the method, we use data simulated at parton-level.
We will ignore any showering, hadronisation, or detector effects.
Future work will extend this method to consider these effects, also incorporating substructure and tracking information.
For now, we treat each final-state particle as a single four-momentum.

\subsection*{States and actions}

The state of the RL environment is defined by the one-to-one mapping between the final-state particle four-momenta and the matrix element partons.
The actions modify the state simply by changing this one-to-one mapping through swapping two particle-parton assignments.
The scattering processes are $2\!\rightarrow\!\text{many}$, so if we have $N$ partons in total, we have $N\!-\!2$ final-state partons.
The action space has dimension $d_{\mathcal{A}}\!=\!1\!+\!(N-2)(N-3)/2$ where $N$ is the total number of partons in the event.
This accounts for each possible parton-swap in the state, along with a `do nothing' action.
We set the maximum number of steps to $T\!=\!N\!-\!3$, since this is the maximum number of swaps needed to re-order any list of size $N\!-\!2$.
Once the agent has performed a parton-swap action during an episode, we mask it for the remainder of the episode so that the same parton-swap action cannot be used twice.

\subsection*{The reward}

Given a state at step $t$, $s_t$, we can compute the matrix element at step $t$, which we call $m_t$.
We use the policy network to compute the action at step $t$, $a_t$, which then allows us to compute the next state $s_{t+1}$ and the next matrix element $m_{t+1}$.
The reward for this action is defined as the log-ratio of the two matrix elements
\begin{equation}
    r_t = \log\frac{m_{t+1}}{m_t}
\end{equation}
such that actions which increase the matrix element more get a larger reward.

\subsection*{Networks}

For the policy network $F_{\text{policy}}(s)$ and target network $F_{\text{target}}(s)$, we use permutation-invariant transformer networks with a linear encoding for the parton four-momenta. 
When the data for a single event is passed to the network, the four-momentum for each parton is passed through a linear embedding network and mapped to a vector of size $d_{\text{model}}\!=\!128$.
The partons, i.e. the nodes in the graph, are assigned an integer ID to define the graph structure.
These integers are passed through a different linear embedding layer and also mapped to a $d_{\text{model}}\!=\!128$ dimensional vector.
While we encode the kinematics and the graph structure separately in the beginning, we then sum these vectors to get a single $d_{\text{model}}$ dimensional vector encoding each parton in the event.
For an event with $N$ partons, we then have a tensor of dimension $(N,d_{\text{model}})$ which is then passed through the self-attention blocks of the transformer network.
The output is again a tensor with shape $(N,d_{\text{model}})$, which we aggregate to a single tensor of dimension $d_{\text{model}}$ by summing over the $N$ parton representations.
This vector is then passed to a dense network that maps the $d_{\text{model}}$ dimensional vector to a vector of size $d_{\mathcal{A}}$.

\subsection*{Training procedure}

In Sec.~\ref{sec:dql} we presented an overview of a generic training procedure for Deep Q-Learning.
In the implementation of our theory-informed model, we have used that same structure and will outline here the specific hyperparameters chosen for the optimisation.
We set the probability of choosing a random action $x_r$ to an initial value of $1.0$ and then have it decay by a factor of $0.999$ at the end of every episode until it reaches $0.5$, after which it remains constant.
However, we only allow these random actions on every second episode.
This means that half of the time the agent is exploring, and half of the time it is allowed to pursue states which may be close to producing the maximum matrix element.
The ensure that the agent is able to learn from a diverse range of state-action combinations, we have used four different replay buffers.
They are split according to the reward values along $(~r_t\!<\!-\!10,~-\!10\!\leq\!r_t\!\leq0,~0\!<\!r_t\!\!\leq\!10,~r_t\!>\!10~)$.
Batches are sampled evenly across all four buffers.
We have not optimised this, and used the same replay buffer setup across all applications presented in this paper.
A summary of the other hyperparameters associated with the training is shown in Tab.~\ref{tab:hyperparameters}.

\begin{table}[h]
    \centering
    \begin{tabular}{l|l}
      hyper-parameter   & value \\ \hline\hline
      \#events   & $10^4$ \\
      \#episodes & $2.5\times10^6$\\
      $\gamma$ & $0.01\!-\!0.5$\\
      buffer size & $10^4$ \\
      initial $x_r$ & $1.0$ \\
      final $x_r$ & $0.5$ \\
      $x_r$ decay & $0.999$ \\
      batch size & $10^3$\\
      optimiser & Adam ($\beta_1\!=\!0.9,~\beta_2\!=\!0.999$)\\
      learning rate & $10^{-5}$ \\
      $d_{\text{model}}$ & $128$ \\
      \#layers & $4$ \\
      \#heads & $4$ \\
      dense net layers & $(128,128,d_{\mathcal{A}})$ \\
      dropout & $0.1$
    \end{tabular}
    \caption{Hyperparameters used in RL training.}
    \label{tab:hyperparameters}
\end{table}

The only hyperparameter that we changed for different applications was the discount factor $\gamma$.
For $t\bar{t}$ and the polarised $WW$+dijet final states, we found that most values worked well and settled for $\gamma\!=\!0.5$.
However, for the higher multiplicity final-states in $t\bar{t}W$ and $t\bar{t}t\bar{t}$ we found that lower values of the discount factor worked much better, settling on $\gamma\!=\!0.01$.
The results were not sensitive to the exact values, and we again did not do an extensive optimisation.

Higher multiplicity final states also required longer training.
The $t\bar{t}$ and $WW$+dijet models converged within a few hundred thousand episodes, while $t\bar{t}W$ took around one million episodes and $t\bar{t}t\bar{t}$ took around two million episodes.
This is to be expected due to the increased complexity in higher multiplicity final states.
In addition to this, the higher-multiplicity final-state networks take additional time to train because there are more steps in each episode.

\begin{figure}[h]
    \centering
    \includegraphics[width=1.0\linewidth]{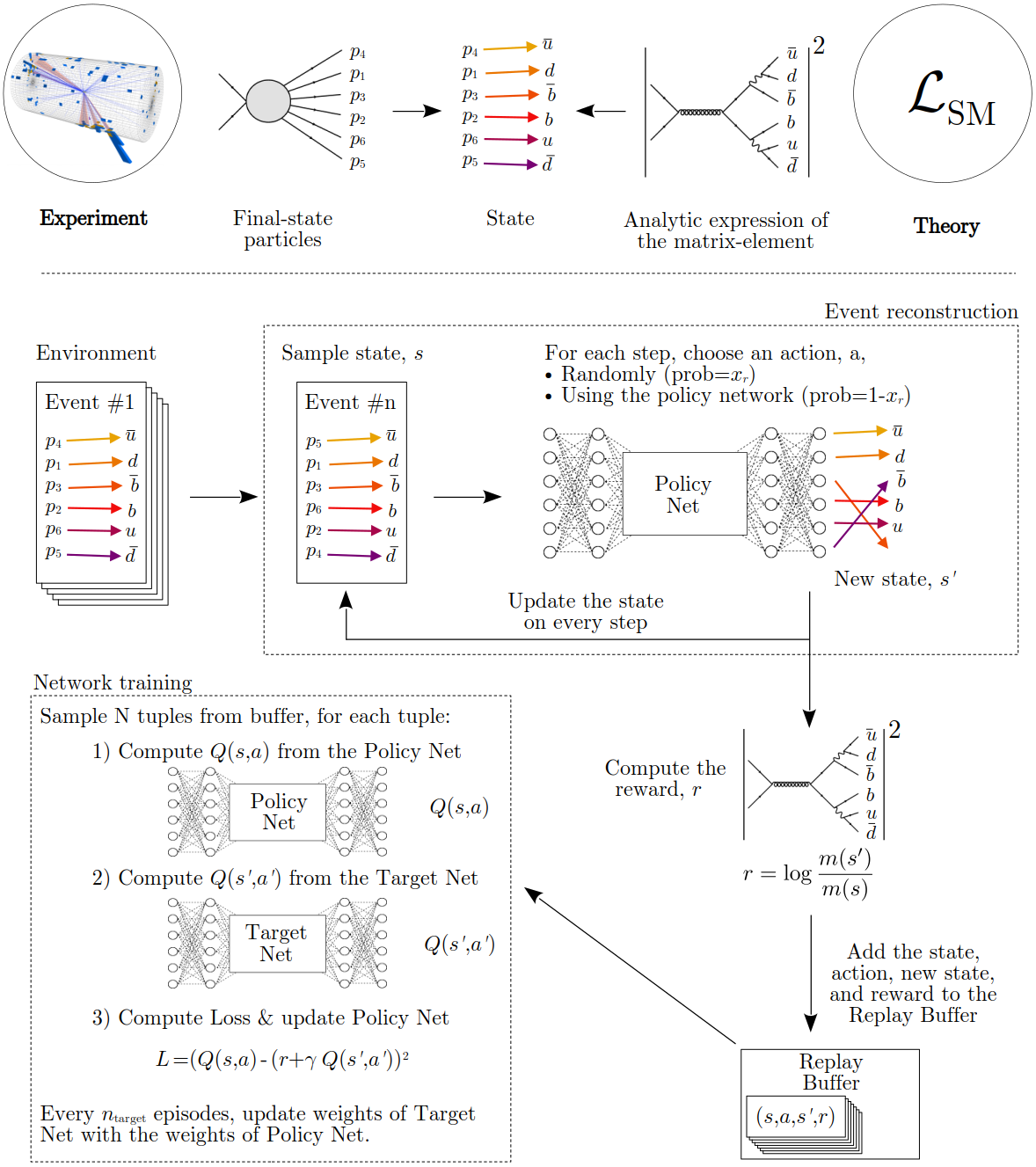}
    \caption{Outline of the Theory-Informed Neural-Network approach to Deep Q-Learning for event reconstruction.}
    \label{fig:rl-schema}
\end{figure}

\section{Theory-informed event reconstruction}
\label{sec:evreco}

In this section, we demonstrate the ability of the method we have outlined to reconstruct events of varying complexity.
We consider three different cases:
\begin{itemize}
    \item[-] $gg\!\rightarrow\! t\bar{t}$
    \item[-] $u\bar{d}\!\rightarrow\! t\bar{t}W$
    \item[-] $gg\!\rightarrow\! t\bar{t}t\bar{t}$
\end{itemize}
where all tops decay to $bW$ and all $W$'s decay hadronically.
Extending to proton-proton initial states does not require any changes to the method whatsoever; we define the processes at parton-level only to demonstrate the method.
We simulated this data, and all data used in further sections of the paper, with MadGraph \cite{Alwall:2014hca} at a centre of mass energy of $13$TeV.
We also use MadGraph to compute the matrix elements used in the rewards during training.
A separate network is trained for each process, although the network architectures are flexible enough to allow for transfer learning between them, and a separate test set of $20$k events is used.
Once the network is trained, we obtain the predicted mapping between final-state particles and matrix element partons by allowing the agent to act on each event in the test set for $N\!-\!3$ steps, where $N$ is the number of partons in the event.

\begin{figure}[h]
    \centering
    \includegraphics[width=0.99\linewidth]{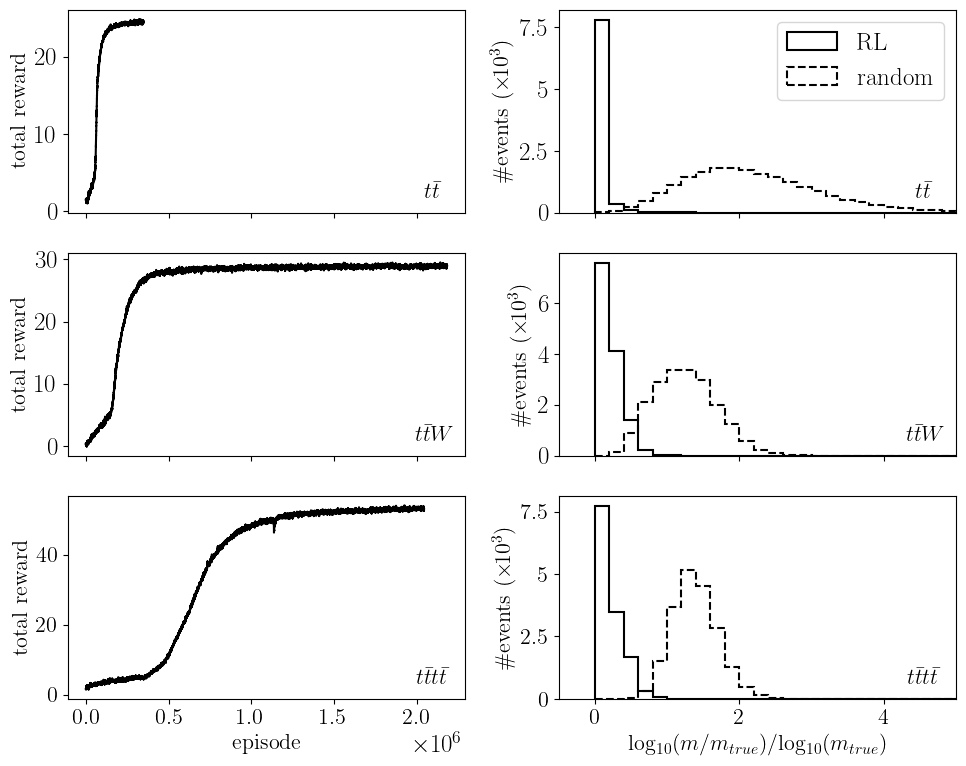}
    \caption{The left plots show the increase in the total reward in each episode over the course of the training.  The histograms on the right show the matrix element after RL vs the true matrix element.}
    \label{fig:perf}
\end{figure}

These three examples span a large range in combinatorial complexity.
The $t\bar{t}$ final-state has only $6$ final-state particles and so there are only $6!\!=\!720$ possible mappings between the final-state particles and the matrix element partons.
For $t\bar{t}W$ this increases to $\sim\!4\times10^4$, and for $t\bar{t}t\bar{t}$ to $\sim\!4\times10^8$.
Impressively, we find that the theory-informed DQL method does a very good job in finding close-to-optimal event reconstructions in all three cases.
In Fig.~\ref{fig:perf} we show both how the total reward over an episode changes throughout the training (left plots), and the relative difference between the matrix element of the predicted mappings and the true matrix element (right plots).
The dashed lines on the right plots show the matrix element for the initial mappings between the final-state particles and the partons, i.e. before the RL method.
We can see that the more difficult the combinatorial problem, the longer the networks take to converge.
In the $t\bar{t}$ case, we only needed to train the network for $\sim 3\!\times\! 10^5$ episodes to reach convergence.
However, in all three cases, we see very good agreement between the matrix elements of the predicted mappings and the true mappings.

\begin{figure}[h]
    \centering
    \includegraphics[width=0.99\linewidth]{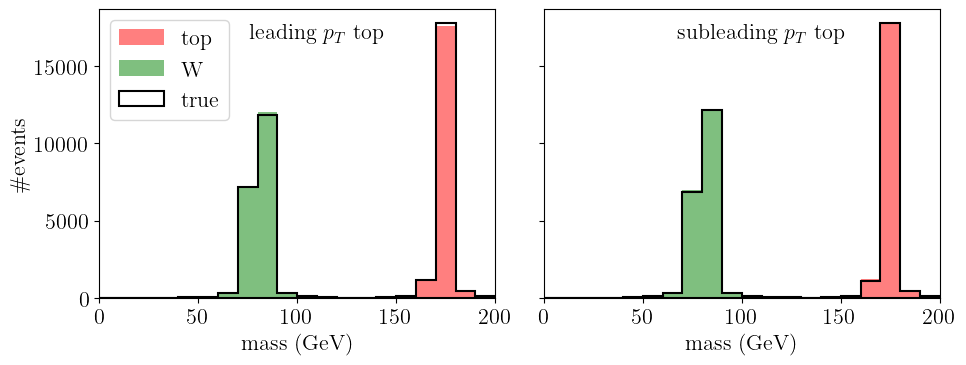}
    \caption{Histogram showing the invariant masses of the reconstructed $W$ boson and top quark masses after applying our RL method on the $t\bar{t}$ data (shaded coloured).  The top is on the left and the anti-top of the right.  The true distributions are shown in black.}
    \label{fig:reco_tt}
\end{figure}

\begin{figure}[h]
    \centering
    \includegraphics[width=0.99\linewidth]{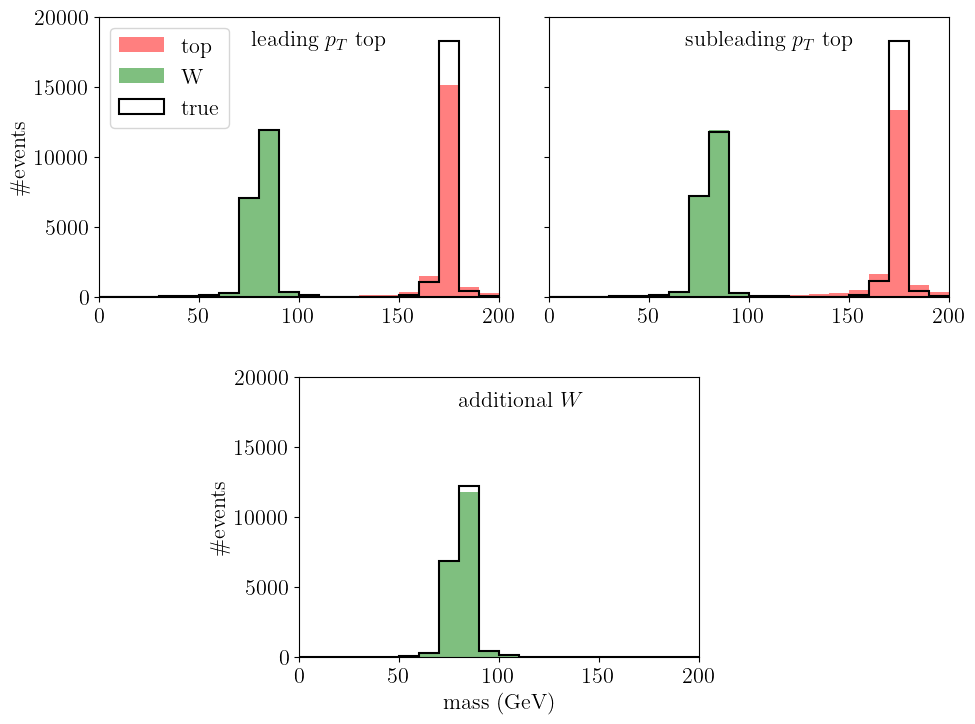}
    \caption{Histogram showing the invariant masses of the reconstructed $W$ boson and top quark masses after applying our RL method on the $t\bar{t}W$ data (shaded coloured).  The top is on the left, the anti-top on the right, and the additional $W$ on the bottom.  The true distributions are shown in black.}
    \label{fig:reco_ttw}
\end{figure}

\begin{figure}[h]
    \centering
    \includegraphics[width=0.99\linewidth]{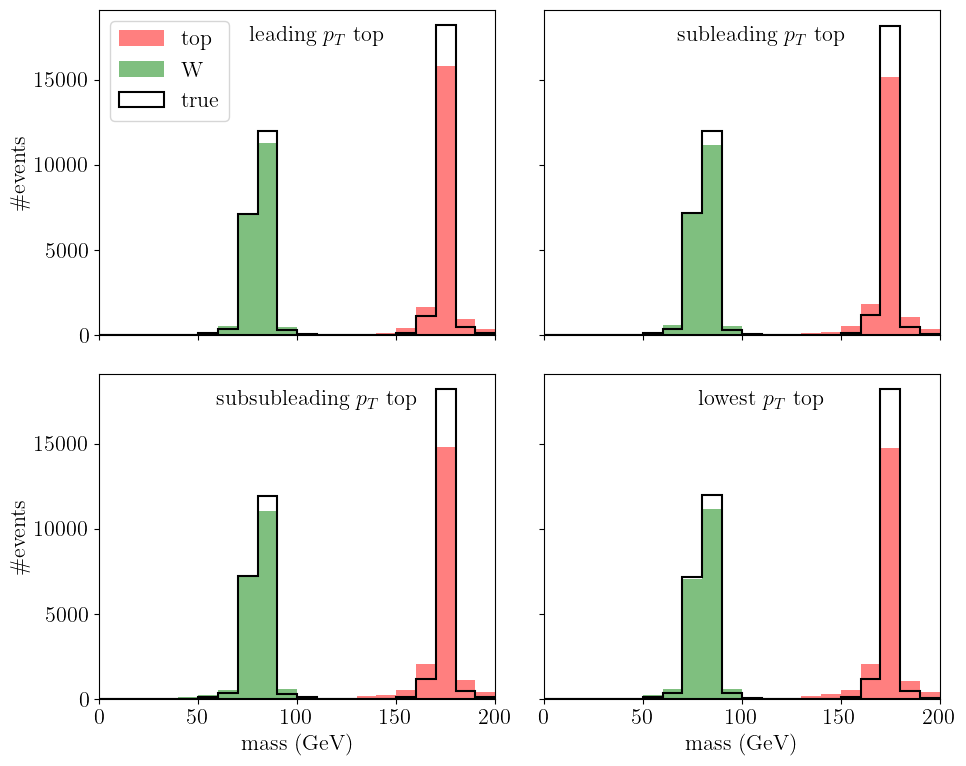}
    \caption{Histogram showing the invariant masses of the reconstructed $W$ boson and top quark masses after applying our RL method on the $t\bar{t}t\bar{t}$ data (shaded coloured).  The tops are on the left, and the anti-tops are on the right.  The true distributions are shown in black.}
    \label{fig:reco_tttt}
\end{figure}

In Fig.~\ref{fig:reco_tt}, \ref{fig:reco_ttw}, and \ref{fig:reco_tttt}, we can see how well the method can reconstruct the mass peaks of the different intermediate particles in each process.
For $t\bar{t}$ we see a near perfect reconstruction, while for $t\bar{t}W$ and $t\bar{t}t\bar{t}$ we see a very good reconstruction with some mis-reconstructed tops.
From inspection of the training process, this seems mostly due to the incorrect matching between the $W$ boson and the $b$ quark.
This is also shown from the fact that the $W$ boson masses are better reconstructed than the top masses for $t\bar{t}W$ and $t\bar{t}t\bar{t}$.
Because the policy network solves the combinatorial particle assignment iteratively, it must first focus on reconstructing at least one $W$ boson within the event before reconstructing any tops.
Forcing the policy network to consider future rewards more, i.e. a better optimisation of the discount parameter $\gamma$, could help improve these results.
Improving these results and increasing the efficiency of the neural networks will be the subject of further work.

\section{Theory-informed classifiers}
\label{sec:ticl}
\subsection*{Longitudinal W-boson tagging}

For the dataset, we have parton-level simulations of $uu\rightarrow W^+W^-uu$ with $ W^+\rightarrow u\bar{d}$ and $W^-\rightarrow \bar{u} d$, for each combination of $W^+W^-$ polarisations (LL,LT,TL,TT).
The aim is to use the theory-informed DQL algorithm to construct a theory-informed tagger for the LL polarised final-states, against the backgrounds (LT,TL,TT).
To do this, we train the algorithm on an unpolarised sample of $W^+W^-uu$ events, such that the reward is a mixture of all four matrix elements.
The actions and, therefore, the optimal particle assignments are computed without knowledge of the $WW$ polarisations.
Once we have the predicted particle assignments for each event ($\tilde{x}_i$), the classification score is computed from the matrix elements for the different polarisations:
\begin{equation}
    \label{eq:cllr}
    L(\tilde{x}_i)=\frac{m_{\text{LL}}(\tilde{x}_i)}{m_{\text{LT}}(\tilde{x}_i) + m_{\text{TL}}(\tilde{x}_i) + m_{\text{TT}}(\tilde{x}_i)}.
\end{equation}
With traditional machine-learning classifiers, the classification score comes directly from a neural network that has been trained on labelled simulated data.
There is very little that can be done to clarify the black-box decision-making process within the network.
With the theory-informed approach here, however, the neural networks only provide the mapping between the final state particles and the partons entering the matrix element.
This neural network output is entirely interpretable; we can study the correlations between different partons and see how well intermediate particle masses are reconstructed.
With the mappings from the neural network output, we then compute the classification score directly using the matrix element.
In the ideal scenario where the neural networks predict the correct mapping from final state particles to the partons, the likelihood ratio defined in Eq.~\ref{eq:cllr} is the optimal classification score.

\begin{figure}[h]
    \centering
    \includegraphics[width=0.99\linewidth]{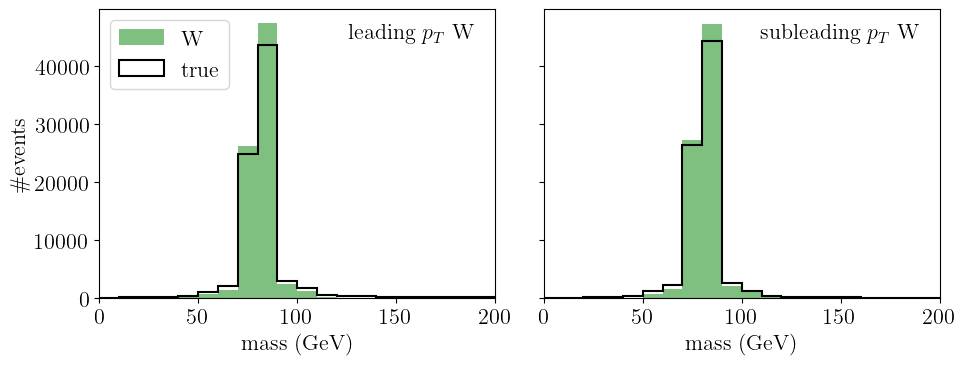}
    \caption{Histogram showing the invariant masses of the reconstructed $W$ bosons on the $W^+W^-uu$ data (shaded coloured).  The true distributions are shown in black.}
    \label{fig:reco_ww}
\end{figure}

\begin{figure}[h]
    \centering
    \includegraphics[width=0.5\linewidth]{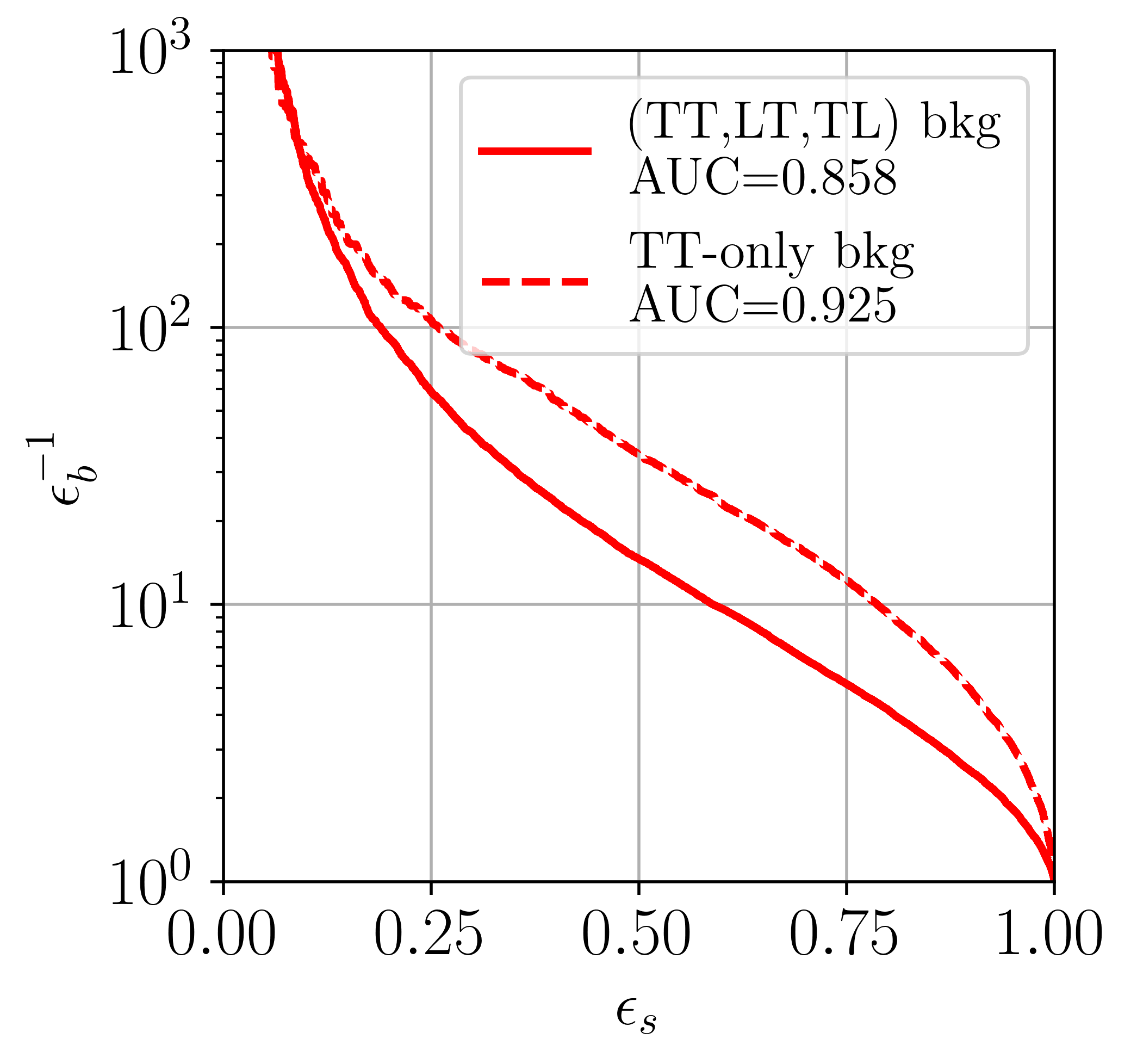}
    \caption{ROC curve showing WW polarisation tagging performance with the signal being the two longitudinally polarised W bosons.   The solid line is the case of the background being either two transverse polarised W bosons or one of each, and the dashed line is when the background is just events with two transverse polarised W bosons.}
    \label{fig:ww-roc}
\end{figure}

We can see the histograms showing the reconstructed masses of the $W$ bosons in Fig.~\ref{fig:reco_ww}, which is in very good agreement with the true mass distribution.
In Fig.~\ref{fig:ww-roc} we see the ROC performance of the classifier for the longitudinal $W$ bosons.
We show ROC curves for both $(LL)$ vs $(TT,LT,TL)$ and $(LL)$ vs $(TT)$, which demonstrates that the method is more easily able to separate the $LL$ final-states from $TT$ alone than it is with all three $(TT,LT,TL)$, as expected.
We can compute the optimal likelihood ratio using the true particle-parton assignments for the different polarisation states, doing so we obtain an AUC of $0.904$ for $(LL)$ vs $(TT,LT,TL)$.
This is the upper limit of what we can achieve with perfect event reconstruction.
However it is difficult to obtain in practice because while training the RL algorithm cannot say a-priori which polarisation states an event has, and so it is trained on the unpolarised sample.

\subsection*{Anomaly-detection}
Lastly we demonstrate how to construct theory-informed anomaly-detection tools using the RL approach described in this paper.
Suppose we have some background process or mixture of background processes for which we can compute the matrix element $m_{BG}$.
We can use the RL algorithm to reconstruct the optimal particle assignments for a dataset that contains a mixture of background events and some anomalous signal events.
With these optimal particle assignments, we can compute the optimal matrix element for each event and use this as a theory-informed anomaly score.
The resulting matrix elements can be interpreted as the likelihoods of these events being produced under the background hypothesis.
Therefore, we define our anomaly score as
\begin{equation}
    \rho(\tilde{x}_i)=\frac{1}{m_{BG}(\tilde{x}_i)}.
\end{equation}
If the signal events are kinematically distinct from the background, their predicted particle assignments should yield a small matrix element and a larger anomaly score.
This method only requires knowledge of the background processes, and is completely agnostic to potential signal models.

\begin{figure}[h]
    \centering
    \includegraphics[width=0.5\linewidth]{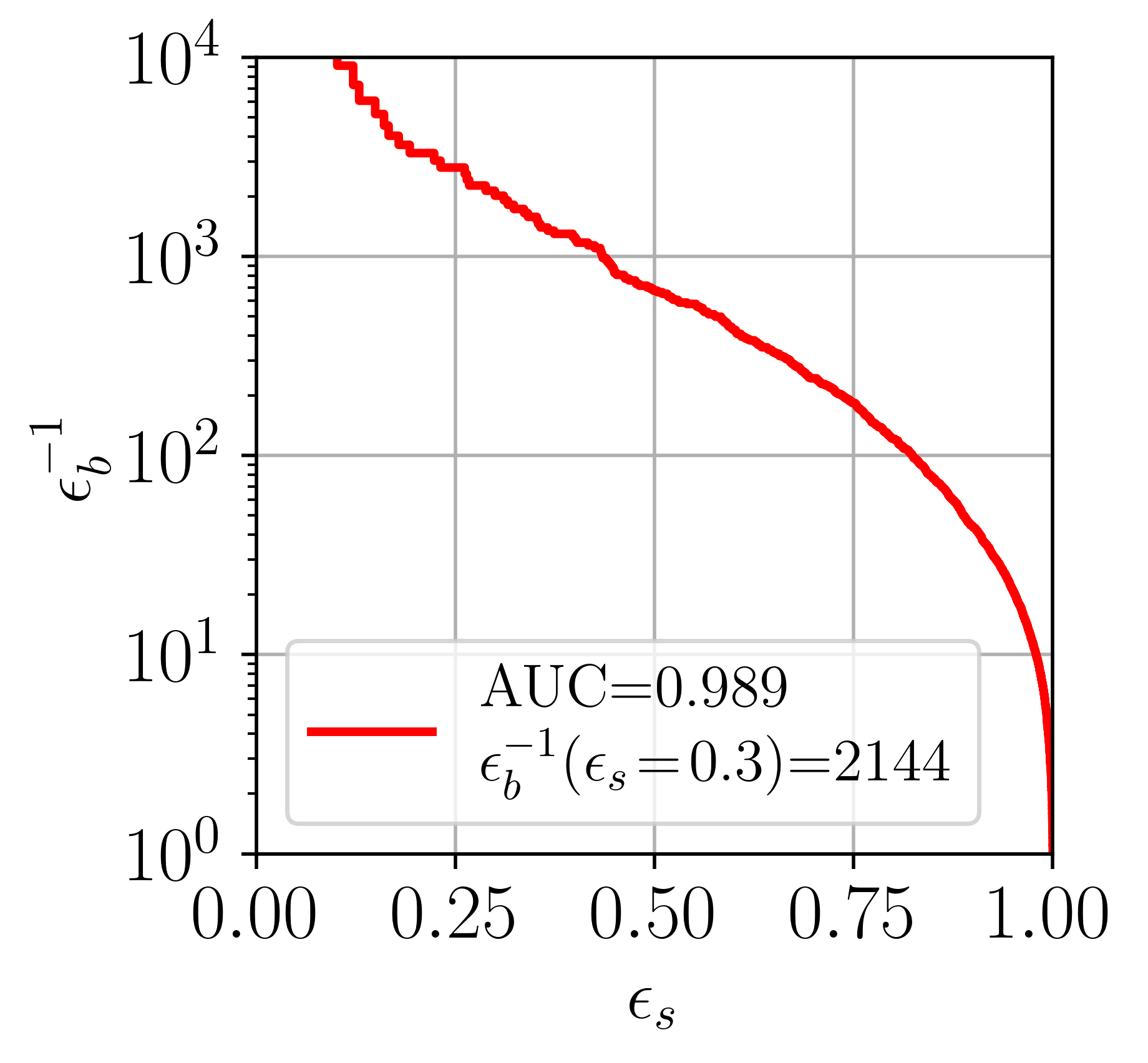}
    \caption{ROC curve showing anomaly-detection performance with a $t\bar{t}$ background and a VBF $WW$ signal, for events with invariant masses in the range $[400,700]$ GeV.}
    \label{fig:anomaly-roc}
\end{figure}

We test this idea with the $t\bar{t}$ and VBF $W^+W^-uu$ datasets used previously, with $t\bar{t}$ being the background and $W^+W^-uu$ being the anomalous signal.
We cut the event samples so that all events have an invariant mass between $[400,700]$ GeV.
The VBF events can typically have much larger invariant masses than the $t\bar{t}$ events, so restricting the invariant mass in this way provides a better test of the anomaly-detection method.
In Fig.~\ref{fig:anomaly-roc} we show the ROC results of this test, demonstrating a powerful anomaly-detection performance, particularly at low signal-efficiencies. 
Being theory-informed, this approach to anomaly-detection does not suffer from the usual implicit biases associated with data-driven anomaly-detection methods such as those constructed with AutoEncoders \cite{kingma2014autoencoding,Farina:2018fyg,Heimel:2018mkt,Blance:2019ibf}.
The AutoEncoders, and other techniques based on density estimation, are coordinate-dependent, meaning that their anomaly scores depend on arbitrary choices associated with data preprocessing \cite{Kasieczka:2022naq}.
Although there do exist physics-inspired preprocessing methods \cite{Piscopo:2019txs,Dillon:2021gag,Dillon:2022tmm,Dillon:2023zac} and techniques to mitigate certain implicit biases \cite{Blance:2019ibf,Dillon:2022mkq,Atkinson:2022uzb,Khosa_2023}, these cannot be removed completely.
On the other hand, techniques based on probabilistic models of the data, such as \cite{Dillon:2019cqt,Dillon:2020quc} may be more sensitive to the masses of intermediate particles in the decay chain, but will still not avail of the full information content in the matrix element.

The theory-informed method that we have shown here is also interpretable.
For any given event we can look at correlations between the partons to check for features that we would expect from background processes, such as intermediate mass peaks.
Data-driven methods typically do not allow such a clear interpretation as they do not provide any information on which final-state particles are assumed to correspond to which partons in the background process.
However some work has been done to overcome the black-box nature of data-driven methods through developing more interpretable techniques \cite{Dillon:2021nxw,Bradshaw:2022qev,Cheng:2025ewj}. 

For comparison, we trained an AutoEncoder on the same anomaly-detection task.
The training data consisted of $100$k $t\bar{t}$ events, with each event containing the energy-ordered 4-vectors of the final state particles flattened to a single $24$ dimensional vector.
The momenta are in units of GeV and are scaled down by a factor of $100$.
The encoder and decoder of the AutoEncoder both have two hidden layers with dimension $64$, and the latent space had dimension $6$.
The network was trained for $500$ epochs, with the loss typically converging at around $200$ epochs.
We used the Mean-Squared-Error loss function, so that the network is trained to compress and reconstruct the training data, while outliers or previously unseen data should be reconstructed poorly.
We use this reconstruction error as the anomaly score.

Testing the AutoEncoder with the $W^+W^-uu$ events being the anomalies, we find that the AutoEncoder is only able to achieve an AUC of $\simeq\!0.645$, with $\epsilon_b^{-1}(\epsilon_s\!=\!0.3)\!\simeq\!8$.
AutoEncoders typically rely on anomalies having features that are very different from the backgrounds, while failing to trigger on more subtle differences.
For example, if we do not cut the event sample to the invariant mass range $[400,700]$ GeV, the performance of the AutoEncoder in identifying the anomalies increases significantly, because the large invariant mass tails in the VBF events leads to larger reconstruction errors.
With the restricted invariant mass however, the AutoEncoder is not sensitive to the same subtler features as the theory-informed method.
It is possible that using more advanced AutoEncoder techniques or making better choices in the data preprocessing will lead to better performance here, but this is beyond the scope of our work.
In addition, constructing Lorentz-equivariant anomaly-detection tools with AutoEncoders has proved complex and challenging \cite{Hao_2023}.
In contrast, the theory-informed anomaly score defined here is naturally Lorentz-invariant, and it is possible to implement Lorentz equivariant transformers for the RL task.
Thus far however, we have only tested this theory-informed method at parton level; future work is required to promote this to a method that fully incorporates the showering, hadronisation, and detector effects. 

\section{Summary and Conclusions}
\label{sec:summary}
We have presented a theory-informed reinforcement-learning framework that reframes the notoriously difficult combinatorial assignment problem in collider physics as a Markov decision process. A Deep Q-Network, guided at every step by the exact tree-level matrix element, learns to manipulate the event graph through successive particle swaps until the global matrix element value is maximised. Because the reward is calculated from first-principles amplitudes rather than from labels, the policy that emerges is inherently interpretable: each node in the final graph corresponds to a well-defined partonic origin, so experimentalists can trace every decision back to quantum-field-theory expressions instead of opaque neural-network weights.

We first tested the event reconstruction capabilities of the method, training the theory-informed networks to reconstruct $t\bar{t}$, $t\bar{t}W$, and $t\bar{t}t\bar{t}$ event topologies using just the four-momenta of the final-state particles.
The networks were able to converge on a solution for $t\bar{t}$ very quickly, and with more training time, they were also able to solve the much more demanding $t\bar{t}W$ and $t\bar{t}t\bar{t}$ topologies.
In each case, we can interpret the output of the network by visualising the correlations between the partons and the masses of intermediate particles in the decay chain.

Going beyond top physics, we turned to $W^{+}W^{-}$ polarisation measurements, a cornerstone test of electroweak symmetry breaking.
Using the unpolarised matrix element for $W^{+}W^{-}$ pair-production in VBF, we trained the network to reconstruct the event topologies.
We then showed how to construct an optimal likelihood ratio classifier for tagging longitudinally polarised pairs using the analytic form of the polarised matrix elements.
The resulting tagger performs well and is fully interpretable.
Lastly, we demonstrate how the same technique can be used to build anomaly-detection scores.
Assuming a $t\bar{t}$ background we treat the $W^{+}W^{-}$ VBF final states as anomalies.
We reconstruct both the background and anomalous events using a network trained to reconstruct only $t\bar{t}$ topologies, and then use the numerical value of the inverse matrix element as the anomaly score.

Because every result rests on exact tree-level amplitudes, the framework requires no labelled data, aligns naturally with likelihood-ratio hypothesis testing, and unifies event reconstruction, tagging, and anomaly searches in a single pipeline.
The machine-learning only provides the mapping between the experiment and theory, while classification and anomaly scores are computed directly from the analytic expression for the matrix element; therefore, the scores respect all physical symmetries in the process.
Future work will incorporate detector effects and systematic variations directly into the reward, extend the action space to handle jet substructure and final-state radiation continuously, explore Lorentz-equivariant network architectures, and include higher-order corrections to reduce theoretical uncertainties. 
These developments promise to sharpen precision Standard-Model measurements and to enhance the discovery potential of the High-Luminosity Large Hadron Collider in channels where complex decay chains currently limit sensitivity.

\subsection*{Acknowledgements}

BMD acknowledges the support of the IPPP through an Associateship.
We are grateful for use of the computing resources from the Northern Ireland High Performance Computing (NI-HPC) service funded by EPSRC (EP/T022175).

\newpage
\printbibliography

\end{document}